\documentclass[fleqn,twoside]{article}
\usepackage{espcrc2}

\usepackage{graphicx}
\usepackage[figuresright]{rotating}

\newcommand{\AmS}{{\protect\the\textfont2
A\kern-.1667em\lower.5ex\hbox{M}\kern-.125emS}}

\def\mee{$\langle m_{\beta\beta} \rangle$}

\def\BBz{$\beta\beta(0\nu)$}
\def\BBt{$\beta\beta(2\nu)$}
\def\BB{$\beta\beta$}
\def\Mz{$|M_{0\nu}|$}

\def\Tt{$T^{2\nu}_{1/2}$}

\def\today{\space\number\day\space\ifcase\month\or January\or February\or
March\or April\or May\or June\or July\or August\or September\or October\or
November\or December\fi\space\number\year}
\title{Neutrino Mass Patterns and Future Double-Beta Decay Experiments}

\author{Steven R. Elliott\address[LANL]{Physics Division, Los Alamos National Laboratory \\ 
MS H803, P-23, Los Alamos, NM 87545, USA}}

\begin{document}

\begin{abstract}
The next generation of double-beta decay experiments have an excellent chance 
of providing data on the neutrino mass pattern. This presentation is a summary of what
is currently known about the mass pattern and expectations from  
experiment. Uncertainties due to the precision in the oscillation 
parameters are not critical to the interpretation of a \BBz\ measurement in terms of the mass
pattern. Even though there is reason for optimism, the matrix 
element uncertainty is still a concern. 
A selected, representative group of the future experiments is discussed. 
\vspace{1pc}
\end{abstract}

\maketitle

\section{Introduction}
Since an introduction to the science of double-beta decay (\BB) was discussed by another speaker
at this meeting\cite{GIU03}, only the critical points necessary for the discussion
in this paper will be summarized here. Reference \cite{ELL02} is a recent review.

The zero-neutrino double-beta decay (\BBz) rate ($\Gamma$), 
is directly related to neutrino mass. $\Gamma$
is proportional to the square of the effective
Majorana neutrino mass (\mee), an easily calculable phase space factor ($G$), and a 
difficult-to-calculate nuclear matrix element (\Mz); 

\begin{equation}
\Gamma = G |M_{0\nu}|^{2} \langle m_{\beta\beta} \rangle^2\mbox{.}
\end{equation}

The value for \mee\ in turn, depends on the values of the individual neutrino mass eigenstates ($m_i$), the
mixing matrix elements ($U_{ei}$) and the Majorana phases ($\alpha_i$);

\begin{equation}
\label{eqn:2}
\langle m_{\beta\beta} \rangle^2 = \left| \sum_i^N  |U_{e i}|^2 e^{i\alpha_i} m_i \right|^2\mbox{.}
\end{equation}

To deduce information on $m_i$ from a measurement  of $\Gamma$ requires values for $|U_{e i}|$ 
and \Mz. In this report, we discuss how the uncertainties in these
factors affect the conclusions on $m_i$. In addition we summarize the proposals for future \BBz\ 
measurements.

\section{OSCILLATIONS AND \BB\ DECAY}

The results of the oscillation experiments provide data on the mixing matrix elements 
and the differences in the squares of the mass eigenvalues ($\delta m_{ij}^2 \equiv m_j^2 - m_i^2$). 
From the atmospheric neutrino
data, we have $\delta m_{23}^{2} \equiv \delta m_{atm}^{2} = 2.0_{-0.7}^{+1.0} 
\times 10^{-3}$ eV$^2$(90\% CL) 
and $\theta_{23} \approx 45$
degrees\cite{SKatm01}. The combined results of the solar neutrino experiments and 
the reactor experiments\cite{AHM03} give 
$\delta m_{12}^{2} \equiv \delta m_{sol}^{2} = 7.1_{-0.6}^{+1.2} \times 10^{-5}$ 
eV$^2$ and $\theta_{12} = 32.5_{-2.3}^{+2.4}$ degrees (68\% CL). (Note that other authors have
found modestly different results. See, {\it e.g.}
Refs. \cite{BAH03,HOL03}.) From reactor experiments, we have a limit
on $\theta_{13} < 9$ degrees\cite{PDB} (68\% CL). If there are only 3 neutrinos, then 
these two $\delta m^2$ values define the mass spectrum given any one of the masses. 
 However, the hierarchy
of the mass spectrum is not yet known. In the convention used in this paper, $\nu_{e}$
 is predominately composed of the mass eigenstate $\nu_1$. The
hierarchy uncertainty can be simply stated: Is $\nu_1$ the lightest
mass eigenstate?

\begin{figure}[htb]
\vspace{9pt}
\includegraphics{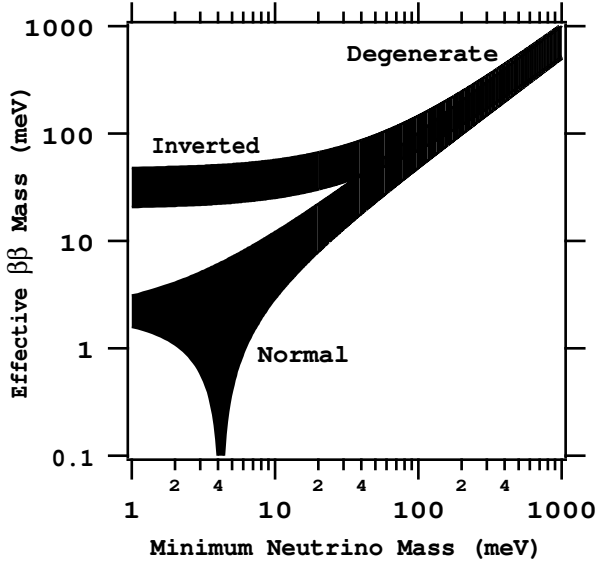}
\caption{The effective Majorana mass as a function of the lightest neutrino mass.}
\label{fig:1}
\end{figure}

The central values of these results
and Eqn. \ref{eqn:2} determine a range of \mee\ values for a given value of $m_1$.
Many authors have done this analysis 
(See, {\it e.g.} Refs. \cite{ELL02,Viss99,Bil99,KPS01,Mat01,Cza01,PPW01,FSV02}.) 
and Fig. \ref{fig:1} shows 
the result of the calculation. The bands indicate the range of possible values
for arbritrary values of the phases. The borders indicate the CP conserving values of the
phases, $e^{i \alpha} = \pm 1$. 

\section{UNCERTAINTIES IN \mee}

The observation of \BBz\ would have profound phyiscs implications regardless of the size of the 
uncertianty in the deduced value of \mee. It would show that neutrinos are massive Majorana
 particles and that the total 
lepton number is not a conserved quantity. But it is interesting to 
ask whether one can use a measurement of \mee\ to discern these two hierarchies. 
At high values of the minimum neutrino mass, the mass spectrum is quasi-degenerate, and the bands
are not resolved. 
At values of the minimum neutrino mass below  $\approx$50 meV, the degenerate band splits into two
bands representing the normal ($m_1$ lightest) and inverted ($m_3$ lightest) hierarchy cases. Figure 
\ref{fig:1} indicates that it would be straight-forward to identify the appropriate band at
these low mass values. However, there are uncertainties 
in the oscillation parameters and the matrix elements that are not represented in the figure. 

One can address this question by comparing the maximum value for the normal 
hierarchy ($\langle m_{\beta\beta} \rangle^{Nor}_{max}$) with the minimum value
for the inverted hierarchy ($\langle m_{\beta\beta} \rangle^{Inv}_{min}$)
that can result from the parameter uncertainies. When the 
lightest neutrino mass is small, one can write expressions for the maximum value for the
normal hierarchy case and the minimum value for the inverted  hierarchy case\cite{PAS03}. When $m_1$
is near zero, $\langle m_{\beta\beta} \rangle^{Nor}_{max}$ occurs for constructive 
interference between the contributions from the
$m_2$ and $m_3$ terms when CP is conserved. 

%\begin{equation}
%\label{eqn:3}
%\langle m_{\beta\beta} \rangle^{Nor}_{max} = \sqrt{\delta m_{sol}^{2}} sin^2\theta_{sol} cos^2\theta_{13} + \sqrt{\delta m_{atm}^{2}} sin^2\theta_{13}
%\end{equation}

\begin{eqnarray}
\label{eqn:3}
\langle m_{\beta\beta} \rangle^{Nor}_{max} & = & \sqrt{\delta m_{sol}^{2}} sin^2\theta_{sol} cos^2\theta_{13}  \nonumber \\
                                           &   & + \sqrt{\delta m_{atm}^{2}} sin^2\theta_{13}  
\end{eqnarray}

From Eqn. \ref{eqn:3}, it is clear that $\langle m_{\beta\beta} \rangle^{Nor}_{max}$ is maximal
when $\theta_{13}$ is maximum, $\theta_{sol}$ is maximum and the $\delta m^2$ are maximum.

$\langle m_{\beta\beta} \rangle^{Inv}_{min}$ is minimal with the
same conditions on $\theta_{13}$ and $\theta_{sol}$, but for a minimum value for $\delta m_{atm}^{2}$.

\begin{equation}
\label{eqn:4}
\langle m_{\beta\beta} \rangle^{Inv}_{min} = \sqrt{\delta m_{atm}^{2}} cos2\theta_{sol} cos^2\theta_{13}
\end{equation}

If we use the appropriate extremum values for the oscillation parameters in Eqns. \ref{eqn:3}
and \ref{eqn:4}, we find 
$\langle m_{\beta\beta} \rangle^{Nor}_{max} \approx (9.1$ meV$)(0.327)(0.976)+(55$ meV$)(0.024) = 4$ meV
and 
$\langle m_{\beta\beta} \rangle^{Inv}_{min} \approx (36$ meV$)(0.345)(0.976) = 12$ meV.
These numbers are sufficienty different, at least when using
these low-CL uncertainty ranges, that it would appear one could discriminate between the two
solutions. 

Since the precision of the oscillation parameters is likely to improve with future experiments, they would
not appear to be the primary concern. 
Even so, how critical each parameter is for this analysis can be determined by propagating its
 uncertainty to the \mee\ uncertainty.
Such a propagation-of-errors analysis is shown in 
Table \ref{table:3} and it is clear that $\theta_{13}$ affects $\langle m_{\beta\beta} \rangle^{Nor}_{max}$ a great deal. 
Its
also clear that $\delta m_{atm}^{2}$ is critical for $\langle m_{\beta\beta} \rangle^{Inv}_{min}$. 
Finally, $\theta_{sol}$  is  important for both. This propagation-of-error analysis, however,
doesn't elucidate the effect that $\theta_{sol}$ also has on Fig. \ref{fig:1}. The value of the
lightest mass for which the  cancellation drives \mee\ to very small values depends
critically on $\theta_{sol}$. But note that as long as the data indicates that the solar mixing is substantially separated
from maximal, the cancellation is possible only over a narrow range of values for the lightest mass.
All in all, the need to interpret $\Gamma$ provides additional motivation to improve the precision 
on $\theta_{13}$, $\theta_{sol}$ and  $\delta m_{atm}^{2}$.

\begin{table*}[htb]
\caption{A summary of the impact on the values for \mee\ in the normal and inverted hierarchies
due to the oscillation parameter uncertainties. For the central values of the parameters,
the nomial values  of $\langle m_{\beta\beta} \rangle^{Nor}_{max}$ 
and $\langle m_{\beta\beta} \rangle^{Inv}_{min}$ are  2.4 meV and 19 meV, respectively. See Ref. 
\cite{PAS03} for a previous similar analysis.}
\label{table:3}
\newcommand{\m}{\hphantom{$-$}}
\newcommand{\cc}[1]{\multicolumn{1}{c}{#1}}
\renewcommand{\tabcolsep}{2pc} % enlarge column spacing
\renewcommand{\arraystretch}{1.2} % enlarge line spacing
\begin{tabular}{@{}cccc@{}}
\hline
Oscillation                  & Parameter        & Range                                    &     Range                             \\  
Parameter                    &   Range          & in  $\langle m_{\beta\beta} \rangle^{Nor}_{max}$ & in $\langle m_{\beta\beta} \rangle^{Inv}_{min}$       \\ 
\hline
$\sqrt{\delta m_{sol}^{2}}$  & 8.1 - 9.1 meV    &	   2.3 - 2.6 meV                       &       N.A.                                     \\
$\sqrt{\delta m_{atm}^{2}}$  & 36 - 55 meV      &	3.2 - 3.7 meV (with $\theta_{13}=9^o$) &       15.2 - 23.2 meV                               \\
$\theta_{sol}$               & 30.1 - 34.9 deg  &	2.1 - 2.7 meV                          &       15.5 - 22.4 meV                          \\
$\theta_{13}$                & 0 - 9 deg        &	2.4 - 3.5 meV                          &       18.6 - 19.0 meV                          \\
\hline
\end{tabular}\\[2pt]
\end{table*}

\Mz\ has been a source of concern for a long time. Typically, an uncertianty of a factor of 2-3
has been assumed for the determination of \mee\ due to \Mz. This uncertainty clearly
dwarfs any uncertainty from the oscillation parameters and thus is the primary
issue. Reference \cite{SUH98} gives an overview of the
calculations, but there has been some recent progress.

$^{76}$Ge is a low-Z isotope relative to most other \BB\ isotopes. Hence it is a good candidate for shell
model calculations of \Mz. The sum over the huge intermediate space of 1$^+$ states was done
by Lanczos moments techniques for the neighboring $^{82}$Se \BB\ isotope \cite{CAU96}. The corresponding 
$^{76}$Ge calculation was done as a series, which is not exactly what is needed. But these calculations
can be improved and new methods\cite{OHS02} might already be able to handle the full-shell model calculation 
for $^{76}$Ge. Such calculations are only as reliable as the input effective interaction and a recent calculation
 by Honma {\it et al.}\cite{HON02} shows that the technology is very
close to doing a Brown-Wildenthal style interaction calculation for Ge. A recent report \cite{ENG03}, however,
demonstrates that single-particle states relatively far from the Fermi level are important for \BBz\
and therefore the required number of states is very large.

The quasiparticle random phase approximation (QRPA) is currently a popular technique used to estimate
\Mz. However, various implementations of QRPA by different authors have produced a spread of results. When
this spread is interpreted as an uncertainty in \Mz, it leads to a factor of 2 uncertainty in \mee. Recently
however, it has been shown\cite{ROD03} that fixing the strength of the particle-particle interaction so
the calculation produces the measured \BBt\ result removes the variablity between the various implementations.
This exciting development will hopefully lead to a better understanding of the source of the spread.

\section{FUTURE EXPERIMENTS}

Table \ref{table:2} summarizes the \BB\ proposals. The target \mee\ sensitivity for the next 
generation of experiments is defined by $\sqrt{\delta m_{atm}^{2}} \approx 45$ meV. At this level, even null
results will have a significant impact on our understanding of the mass spectrum if neutrinos are Majorana
particles. This is especially true if the result is 
coupled with a kinematic measurement of neutrino mass from either tritium beta decay\cite{Katrin}
or cosmology. (See {\it e.g.} Ref. \cite{wmap}.) To accomplish this goal 
requires approximately 1 ton of isotope.

Of the projects listed in Table \ref{table:2}, five are especially worthy
of extra notice in my opinion. These five, CUORE, EXO, GENIUS, Majorana and MOON, have designs that meet the technical 
requirements for the 45-meV goal and, as a group, 
span the various
techniques used for the study of \BB. Since
 CUORE was described by another speaker\cite{GIU03}, I will only 
describe the other four projects.

\begin{table*}[htb]
\caption{A summary of the double-beta decay proposals\cite{ELL02}. The quoted sensitivities are 
those quoted by the proposers but scaled for 5 years
of run time. These sensitivities should be used carefully as they depend on 
background estimates for experiments that don't yet exist.}
\label{table:2}
\newcommand{\m}{\hphantom{$-$}}
\newcommand{\cc}[1]{\multicolumn{1}{c}{#1}}
\renewcommand{\arraystretch}{1.2} % enlarge line spacing
\begin{tabular}{@{}lclc@{}}
\hline
&            &                                                                   &  Sensitivity to         \\  
Experiment                 &	Source     & Detector Description                                           & $T_{1/2}^{0\nu}$ (y)    \\ 
\hline
CAMEO\cite{BEL01}          &$^{116}$Cd  &	1 t  CdWO$_4$ crystals in liq. scint.                           & $1 \times 10^{27}$   \\
CANDLES\cite{KIS01}        &$^{48}$Ca   &	several tons of CaF$_2$ crystals in liq. scint.                 & $1 \times 10^{26}$      \\
COBRA\cite{ZUB01}          &$^{130}$Te  &	10 kg CdTe semiconductors                                       & $1 \times 10^{24}$   \\
CUORE\cite{AVI01}          &$^{130}$Te  &	750 kg TeO$_2$ bolometers                                       & $2 \times 10^{26}$    \\
DCBA\cite{ISH00}           &$^{150}$Nd  &	20 kg $^{enr}$Nd layers between tracking chambers               & $2 \times 10^{25}$     \\
EXO\cite{EXO00}            &$^{136}$Xe  &	1 t $^{enr}$Xe TPC (gas or liquid)                              & $8 \times 10^{26}$    \\
GEM\cite{ZDE01}            &$^{76}$Ge   &	1 t $^{enr}$Ge diodes in liq. nitrogen                          & $7 \times 10^{27}$      \\
GENIUS\cite{KLA01b}        &$^{76}$Ge   &	1 t 86\% $^{enr}$Ge diodes in liq. nitrogen                     & $1 \times 10^{28}$    \\
GSO\cite{DAN01,WANGS01}    &$^{160}$Gd  &	2 t Gd$_2$SiO$_5$:Ce crystal scint. in liq. scint.              & $2 \times 10^{26}$   \\
Majorana\cite{MAJ01}       &$^{76}$Ge   &	0.5 t 86\% segmented $^{enr}$Ge diodes                          & $3 \times 10^{27}$   \\  
MOON\cite{EJI00}           &$^{100}$Mo  &	34 t $^{nat}$Mo sheets between plastic scint.                   & $1 \times 10^{27}$   \\   
NEMO 3\cite{NEMO3}         &$^{100}$Mo  &	10 kg of \BBz\ isotope (7 kg Mo) with tracking                  & $4 \times 10^{24}$    \\   
Xe\cite{CAC01}             &$^{136}$Xe  &   1.56 t of $^{enr}$Xe in liq. scint.                               & $5 \times 10^{26}$    \\
XMASS\cite{XMASS}          &$^{136}$Xe  &   10 t of liq. Xe                                                  & $3 \times 10^{26}$    \\ 
\hline
\end{tabular}\\[2pt]
\end{table*}

\subsection{EXO}
The Enriched Xenon Observatory (EXO)\cite{EXO00} 
proposes to use up to 10 t of 60-80\% enriched $^{136}$Xe.
The unique aspect of this proposal is the plan to detect 
the $^{136}$Ba daughter ion correlated with the
decay. If the technique is perfected, it would eliminate 
all background except that associated with \BBt. 
The real-time optical detection of the daughter Ba ion, initially suggested in Ref. \cite{MOE91},  might 
be possible if the ion can be localized and probed with lasers. The spectroscopy has been used for 
Ba$^+$ ions in atom traps. However, the additional technology to 
detect single Ba ions in a condensed medium 
or to extract single Ba ions from a condensed medium and trap them must be 
demonstrated for this application. 

The EXO plan is to use Liquid Xe (LXe) 
scintillator. The LXe concept has the advantage of being much smaller than a gaseous TPC due to the high density
 of LXe. Once the Ba ion is localized via its scintillation 
and ionization, it might be extracted 
via a cold finger electrode coated in frozen Xe (M. Vient, unpublished observation, 1991). 
The ion is electrostatically attracted to
 the cold finger which later can be heated to evaporate the Xe and release the Ba ion into a radio 
frequency quadrupole trap. At that point, the Ba$^{++}$ is neutralized to Ba$^+$, laser cooled and
 optically detected. 

The collaboration has recently performed experiments to optimize the energy resolution\cite{EXOResol}.
By measuring both scintillation light and ionization simultaneously, they have achieved energy resolution
sufficient for the experiment. Tests to determine the viability of the Ba
 extraction  process are also being performed. 
The EXO collaboration has received funding to proceed with a 200-kg enriched Xe
detector without Ba tagging. This initial prototype will operate at the Waste Isolation Pilot Plant
(WIPP) in southern New Mexico.

\subsection{GENIUS}
The progress and understanding of Ge detectors has been developed over more than 30 years 
of experience. The potential of these detectors lie in their great energy resolution, ease of 
operation, and the extensive experience relating to the reduction of backgrounds. 
This potential is not yet exhausted as is evidenced by the GENIUS and Majorana proposals .

The GENIUS (GErmanium NItrogen Underground Setup)\cite{KLA01b}
 proposal has evolved from the Heidelberg-Moscow (HM) experiment.
 The driving design principle behind this proposed Ge detector array experiment is the 
evidence that the dominant background in the HM experiment was due to radioactivity external
 to the Ge. (The reader should contrast this with the motivation for the design of the 
Majorana proposal described below.) An array of 2.5-kg, p-type Ge crystals would be operated
 ''naked" within a large liquid nitrogen (LN) bath. By using naked crystals, the external 
activity would be moved to outside the LN region. 
 Due to its low stopping power, roughly 12 m of LN is required to shield the crystals from 
the ambient $\gamma$-ray flux at the intended experimental site at 
Gran Sasso. A test of the naked operation of a crystal in a 50 l
 dewar has been successful\cite{KLA98,BAU99}. 

The proposal anticipates an energy resolution of $\approx 6$ keV FWHM (0.3\%) and a threshold of 11 keV. The value of
 this low threshold is set by x rays from cosmogenic activities. Using 1 t of 86\% enriched 
Ge detectors, the target mass is large enough for dark matter
 studies. In fact a 10-kg $^{nat}$Ge proof-of-principle experiment for dark
 matter studies has begun at Gran Sasso\cite{GTF}.

\subsection{Majorana} The Majorana Collaboration 
plans to use $\approx$ 500, 86\% enriched, segmented Ge crystals for a total of 500 kg of
 detector\cite{MAJ01,MAJ03}. The cryostat would be formed from very 
 pure electroformed Cu\cite{BRO90}.
 Their analysis indicated that $^{68}$Ge contained within the Ge detectors
 was the limiting background for their \BBz\ search. (Contrast this with the GENIUS approach
described above.) The proposal's design therefore 
emphasizes segmentation and pulse shape discrimination to reject this background. 
$^{68}$Ge produces a background that deposits energy at multiple sites in the detector.
 In contrast, a \BBz\ event will have a localized 
energy deposit. Segmentation of the crystals permits a veto of such events. Furthermore,
 distinct ionization events will have a different pulse shape than a localized event. 
Thus pulse shape analysis can also reject background.   
 
 The collaboration is fielding a close-packed array of 18 Ge detectors, 16 of which will share
 a lone cryostat. This prototype, called MEGA\cite{MEGA}, will demonstrate the cryogenic
 cooling design for multiple crystals in a single low-background cryostat. It will also
 permit a study of crystal-to-crystal coincidence suppression of backgrounds for \BB\ and 
 dark matter. Finally, the operation of MEGA at WIPP will provide an excellent material
 screening facility in addition to a very sensitive apparatus for studying \BB\ to excited states.
 These later experiments will be conducted by placing samples within the MEGA detector 
 arrangement. The high efficiency for $\gamma$-ray detection will provide the sensitivity
 for observing the two $\gamma$ rays from the excited state relaxation.
 
 In addition, the collaboration is studying the performance of several segmented detector configurations.
 This program, call SEGA\cite{SEGA}, will establish the background rejection capabilities of
 segmented detectors experimentally and confirm the previous Monte Carlo studies. Pulse shape
 discrimination tends to identify multiple energy deposits along the radial direction in crystals
 whereas the segmentation tends to identify multiple deposits axially and azimuthally. The 
 SEGA program will measure the orthogonality of these cuts. There are 3 segmented geometries
 that will be studied; (1) a custom designed, isotopically enriched, 12-segment detector (2)
 a stock-item commercially available Clover (TM) detector from Canberra and (3) a many-segmented
 detector originally obtained for studies for the GRETA project\cite{GRETINA}. The 12-segmented enriched
 detector will also operate at WIPP, both as a stand-alone unit and also as one of the detectors
 comprising the MEGA apparatus.

\subsection{MOON}

The MOON (Mo Observatory Of Neutrinos) proposal\cite{EJI00} plans to use $^{100}$Mo as a \BBz\ source
 and as a target for solar neutrinos. This dual purpose and a sensitivity to low-energy supernova electron
neutrinos\cite{EJI01}
 make it an enticing idea. $^{100}$Mo has a high Q-value (3.034 MeV), which results in a large
phase space factor and places the \BBz\ peak region well above most radioactive backgrounds. It also 
has hints of a favorable
 \Mz\ but unfortunately it has a relatively short \Tt. The experiment will make energy and
angular correlation studies of \BB\ to select \BBz\ events and to
reject backgrounds. The planned MOON configuration is a supermodule of scintillator and Mo ensembles. One
option is a module of plastic fiber scintillators with thin (20 mg/cm$^2$)
layers of Mo, arranged to achieve a position resolution 
comparable to the fiber diameter (2-3 mm).  

The project needs Mo and scintillator radioactive impurity levels less than 1 mBq/ton.
 This can be achieved by carbonyl chemistry for Mo and plastics can be produced cleanly. 
The simulations of the scintillator-film sandwich design indicate that the energy resolution for the \BBz\ peak will be
 $\approx$5\% FWHM, which is at the upper end of the range of feasibility for a sub 50 meV \mee\ experiment.
Metal-loaded liquid scintillator and bolometer options are also being considered. 
The bolometer option would remove the resolution concerns.
Use of enriched  $^{100}$Mo is feasible, as it can be enriched by either gas centrifuge or laser ionization
separation. Enrichment would reduce
the total volume of the apparatus resulting in a lower
internal radioactivity contribution to the background by an order of magnitude.

\subsection{OTHER PROPOSALS}

There are too many additional proposals in Table \ref{table:2} for detailed description but 
I mention the remaining ones here. The CAMEO proposal\cite{BEL01}
 would use 1000 kg of scintillating $^{116}$CdWO$_4$
crystals situated within the Borexino apparatus. The Borexino liquid scintillator would provide
shielding from external radioactivity and light piping of crystal events to 
the photomultiplier tube (PMT) array surrounding
the system.  Similarly, 
the CANDLES proposal\cite{KIS01} (CAlcium floride for study of Neutrino and Dark matter by Low Energy Spectrometer) plans to 
immerse CaF$_2$ in liquid scintillator. The scintillation light from the $\beta\beta$ of $^{48}$Ca will be detected
via PMTs. Two groups\cite{DAN01,WANGS01} have been studying the use of  GSO crystals (Gd$_2$SiO$_5$:Ce) for the study of
\BB\ in $^{160}$Gd. 

COBRA (CdTe O neutrino double Beta Research Apparatus)\cite{ZUB01} 
would use CdTe or CdZnTe semiconductors to search for \BBz\ in either Cd or Te. 1600 1-cm$^3$ 
crystals would provide 10 kg of material. GEM is a proposal\cite{ZDE01} that is very similar to that
of GENIUS. However, much of the LN shielding would be  replaced with pure
water.

The Drift Chamber Beta-ray Analyzer (DCBA) proposal\cite{ISH00} is for a three-dimensional tracking chamber in a 
uniform magnetic field. 
Thin plates of Nd would form the source. The series of NEMO experiments is progressing with NEMO-3\cite{NEMO3}
that began operation in 2002. The detector 
contains a source foil enclosed between tracking chambers that is itself enclosed within a scintillator
array. NEMO-3 can contain a total of 10 kg of source and plans to operate 
with several different isotopes,
but with $^{100}$Mo being the most massive at 7 kg. The collaboration is also discussing 
the possibility
of building a 100-kg experiment that would be called NEMO-4.

There are two additional groups proposing to use $^{136}$Xe to study \BBz. Caccianiga and Giammarchi\cite{CAC01} propose to 
dissolve 1.56 t of enriched Xe in liquid scintillator. The XMASS\cite{XMASS} collaboration proposes to use
10 t of liquid xenon for solar neutrino studies. The detector would have sensitivity to \BBz.

\section{CONCLUSIONS}
This is a very exciting time for \BB. The next generation of experiments will be sensitive to a mass region where
neutrino masses are known to exist. As a result, even null experiments will have an impact on our understanding of 
the mass spectrum. The subtleies associated with the uncertainties in $|U_{e i}|$ 
and \Mz are of secondary importance. If \BBz\ is observed, the physics learned will be 
revolutionary as it would establish the neutrino as a massive
Majorana particle.

\section{ACKNOWLEDGMENTS}
I wish to thank Mark Boulay, Harry Miley, and Petr Vogel for critical readings of this manuscript. 
I thank Wick Haxton for a discussion
of the shell model situation. This research was sponsored in part
by Los Alamos Laboratory-Directed Research and Development.

\end{document}